\documentclass[runningheads]{llncs}
\usepackage{graphicx}
\usepackage{wrapfig}
\usepackage[
  separate-uncertainty = true,
  multi-part-units = repeat
]{siunitx}
\usepackage{mathtools}
\usepackage{graphicx}
\usepackage{courier}
\usepackage{smartdiagram}
\usepackage{epstopdf}
\usepackage{color}
\usepackage{csquotes}
\usepackage{caption} 
\usepackage{mathtools}
\usepackage{amsfonts}
\usepackage{amssymb}
\usepackage[subnum]{cases}
\usepackage{booktabs} 
\usepackage{pifont}
\usepackage{graphics}
\usepackage{amssymb}
\usepackage{pifont}

\graphicspath{ {./Images/} }

\usepackage {tikz}
\usetikzlibrary {positioning}
\definecolor {processblue}{cmyk}{0.96,0,0,0}
\usepackage{lipsum,adjustbox}

\usepackage[subnum]{cases}
\usepackage{color,soul}

\usepackage{algorithm,algorithmicx,algpseudocode}

\usepackage{subcaption}

\usepackage{relsize}

\usepackage{footnote}
\makesavenoteenv{tabular}

\usepackage{multirow}

\usepackage{hyperref}
\usepackage{cleveref}[2012/02/15]

\crefformat{footnote}{#2\footnotemark[#1]#3}

\begin{document}
\title{Line Hypergraph Convolution Network: Applying Graph Convolution for Hypergraphs}
\titlerunning{Line Hypergraph Convolution Network}
%

\author{Sambaran Bandyopadhyay\inst{1,2} \and
Kishalay Das\inst{2} \and
M. Narasimha Murty\inst{2}}

\authorrunning{S. Bandyopadhyay et al.}

\institute{IBM Research,
\and
Indian Institute of Science, Bangalore\\
\email{sambband@in.ibm.com, \{kishalaydas,mnm\}@iisc.ac.in}}

\maketitle              
\begin{abstract}
Network representation learning and node classification in graphs got significant attention due to the invent of different types graph neural networks. Graph convolution network (GCN) is a popular semi-supervised technique which aggregates attributes within the neighborhood of each node. Conventional GCNs can be applied to simple graphs where each edge connects only two nodes. But many modern days applications need to model high order relationships in a graph. Hypergraphs are effective data types to handle such complex relationships. In this paper, we propose a novel technique to apply graph convolution on hypergraphs with variable hyperedge sizes. We use the classical concept of line graph of a hypergraph for the first time in the hypergraph learning literature. Then we propose to use graph convolution on the line graph of a hypergraph. Experimental analysis on multiple real world network datasets shows the merit of our approach compared to state-of-the-arts.

\end{abstract}
\section{Introduction}\label{sec:intro}
Graph representation learning got remarkable attention in the last few years. Performance on semi-supervised tasks such as node classification in a graph has improved significantly due to invent of different types of graph neural networks (GNNs) \cite{niepert2016learning,velivckovic2017graph}. Graph convolution network (GCN) \cite{kipf2016variational} is one of the most popular and efficient neural network structure that aggregates the transformed attributes over the neighborhood of a node. The transformation matrix is learned in a semi-supervised way by minimizing the cross entropy loss of predicting the labels of a subset of nodes. Graph convolution network \cite{kipf2016variational} is an fast approximation of spectral graph convolutions \cite{hammond2011wavelets}. There are different variants of graph convolution present in the literature such as inductive version of GCN with a different aggregation function \cite{hamilton2017inductive}, primal-dual GCN \cite{monti2018dual}, FastGCN \cite{chen2018fastgcn}, etc.

Most of the existing graph convolution approaches are suited for \textit{simple graphs}, i.e., when the relationship between the nodes are pairwise. In such a graph, each edge connects only two vertices (or nodes). However, real life interactions among the entities are more complex in nature and relationships can be of high-order (beyond pairwise connections). For example, when four authors write a research papers together, it is not necessary that any two of them are connected directly \cite{han2009understanding}. They are still coauthors because of some other coauthor in the paper who is strongly connected to both of them. Thus, representing such a co-authorship network by a simple graph may not be suitable. Hypergraphs \cite{bretto2013hypergraph} are introduced to model such complex relationships among the real world entities in a graph structure. In a hypergraph, each edge may connect more than two vertices. So an edge is essentially denoted by a subset of nodes, rather than just a pair. Computation on hypergraphs are more expensive and also complicated. But due to their power to capture real world interactions, it is important to design learning algorithms for hypergraph representation. 

Network analysis community shows interest in different applications of hypergraph mining \cite{chen2015multimodal,zhang2017re}.
There exist different ways to transform a hypergraph to a simple graph such as via clique expansion and star expansion \cite{zhou2007learning,agarwal2006higher,pu2012hypergraph}. Clique expansion creates cliques by connecting any pair of nodes belonging to the same hyperedge in the transformed simple graph. The star expansion creates a bipartite graph by creating a new node for each hyperedge. Typically, the analysis on hypergraph is done by performing the downstream mining tasks on the transformed simple graph.
There are also tensor based approaches \cite{kolda2009tensor} available to deal with hypergraphs, but typically they assume the size of all the hyperedges in the hypergraph to be the same (i.e., uniform hypergraph\footnote{A k-uniform hypergraph is a hypergraph such that all its hyperedges have size k.}) \cite{shashua2006multi,bulo2009game}.
Very recently, few graph neural networks based approaches are proposed for hypergraphs \cite{feng2019hypergraph,jiang2019dynamic,yadati2019hypergcn}. They use clique expansion or hypergraph Laplacian to use the convolution on the hypergraph. Our work is towards this direction. Following are the novel contributions we make in this paper.
\begin{itemize}
    \item We propose a novel approach of applying graph convolution on hypergraphs. We refer our proposed algorithm as LHCN (\textbf{L}ine \textbf{H}ypergraph \textbf{C}onvolution \textbf{N}etwork). To the best of our knowledge, we use the classical concept of \textit{line graph of a hypergraph} first time in the \textit{hypergraph learning literature}. We map the hypergraph to a weighted and attributed line graph and learn the graph convolution on this line graph. We also propose a reverse mapping to get the labels of the nodes in the hypergraph. The proposed algorithm can work with any hypergraphs, even with the non-uniform ones. 
    \item We conduct thorough experimentation on popular node classification datasets. Experimental results show that the performance of LHCN is as per, and often improve the state-of-the-arts in hypergraph neural networks. We make the source code publicly available at \url{https://bit.ly/2qNmbRn} to ease the reproducibility of the results.
\end{itemize}

\section{Related Work}\label{sec:related}
To give a brief but comprehensive idea about the existing literature, we discuss some of the prominent works in the following three domains.

\textbf{Network Embedding and Graph Neural Networks}: A detailed survey on network representation learning and graph neural networks can be found in \cite{wu2019comprehensive,hamilton2017representation}. Different types of semi-supervised graph neural networks exist in the literature for node representation and classification \cite{niepert2016learning,chen2018fastgcn}.
\cite{kipf2016semi} proposes a version of graph convolution network (GCN) which learns a weighted mean of neighbor node attributes to find the embedding of a node by minimizing the cross entropy loss for node classification. \cite{hamilton2017inductive} proposes GraphSAGE which employs different types of aggregation methods for GCN and extends it for inductive node classification. \cite{velivckovic2017graph} proposes GAT which uses attention mechanism to learn the importance of a node to determine the label of another node in the neighborhood of it in the graph convolution framework.
Recently, a GCN based unsupervised approach (DGI) is proposed \cite{velivckovic2018deep} by maximizing mutual information between patch and high-level summaries of a graph.

\textbf{Learning on Hypergraphs}: As mentioned in Section \ref{sec:intro}, many of the existing analysis on hypergraph first transform the hypergraph to a simple graph by clique expansion or star expansion \cite{agarwal2006higher,zhou2007learning,pu2012hypergraph}, and then do the mining on the simple graph. Conventional analysis on simple graphs are done on the adjacency matrix as it captures the graph structure well. Similarly, in the hypergraph domain, higher order matrices, called tensors \cite{borisenko1968vector,kolda2009tensor}, are used for multiple computations. A non-negative tensor factorization approach is proposed in \cite{shashua2006multi} for clustering a dataset having complex relations (beyond pairwise) in the form of a hypergraph.
A hypergraph clustering approach by casting it into a non-cooperative multi-player game is proposed in \cite{bulo2009game}. The main disadvantage of tensor based approaches are that, mostly they assume the hypergraph to be uniform, i.e., all the hyper edges are of equal size. Link prediction in hypergraphs is also studied in the literature \cite{chen2015multimodal,sharma2014predicting}. Submodular hypergraphs are introduced in \cite{li2018submodular}, which arise in clustering applications in which higher-order structures carry relevant information. Recently, a hypergraph based active learning scheme is proposed in \cite{chien2019hs}, which allows one to ask both pointwise queries and pairwise queries.

\textbf{Hypergraph Neural Networks}: Application of graph neural networks for hypergraphs is still a new area of research. To the best of our knowledge, there are only three prominent works in this. A hypergraph neural network (HGNN) is proposed in \cite{feng2019hypergraph} which applies convolution on the hypergraph Laplacian. From Eq. 10 of \cite{feng2019hypergraph}, the framework boils down to the application of graph convolution (as proposed in \cite{kipf2016semi}) on a weighted clique expansion of the hypergraph, where the weights of the edges of a clique is determined by the weight and degree of the corresponding hyperedge. Assuming the initial hypergraph structure is weak, dynamic hypergraph neural network \cite{jiang2019dynamic} is proposed by extending the idea of HGNN, where a dynamic hypergraph construction module is added to dynamically update hypergraph structure on each layer. Very recently, HyperGCN is proposed in \cite{yadati2019hypergcn}, where the authors use the maximum distance of two nodes (in the embedding space) in a hyperedge as a regularizer. They use the hypergraph Laplacian proposed in \cite{chan2018spectral} to transform a hypergraph into a simple graph where each hyperedge is represented by a simple edge and the edge weight is proportional to the maximum distance between any pair of nodes in the hypergraph. Then they perform GCN on this simple graph structure.

Our proposed approach belongs to the class of hypergraph neural networks, where we invent a novel method to apply graph convolution on the hypergraphs.

\section{Problem Statement and Notations Used}\label{sec:prob}
We consider an undirected hypergraph $\mathcal{H} = (V,E)$. Here $V$ is the set of (hyper)nodes with $|V|=n$ and $E$ is the set of hyperedges, with $|E|=m$. A hyperedge $e$ connects a subset of nodes. For example, if the hyperedge $e$ connects $v_1$, $v_2$ and $v_3$, it is denoted as $e=\{v_1,v_2,v_3\}$.
A hypergraph is often represented by an incidence matrix $H \in \mathbb{R}^{n \times m}$, where $H(i,j) = 1$ if the hyperedge $j$ contains the node $i$, and $H(i,j) = 0$ otherwise.
We also assume the hypergraph to be attributed, i.e., each node $v \in V$ is associated with a $d$ dimensional feature vector $x_v \in \mathbb{R}^d$ and this forms a feature matrix $X \in \mathbb{R}^{n \times d}$. We aim to propose a novel algorithm to apply graph convolution for the hypergraph which can be used to classify the nodes of the hypergraph. So we assume to have a training set $V^l \subset V$ where for each node $v \in V^l$, we know the label $l_v \in \mathcal{L}$ of it. Here, $\mathcal{L}$ is the set of labels. For example, $\mathcal{L}=\{-1,+1\}$ for a binary classification. Our goal is to learn a function $f : V \mapsto \mathcal{L}$ which can output label of each unlabelled node $v \in V^u = V \setminus V^l$. The desired algorithm to learn such a function $f$ should be able to use both the hypergraph structure, along with the node attributes.

\section{Solution Approach: LHCN}\label{sec:soln}
In this section, we discuss each stage of the proposed algorithm LHCN. We also give the necessary background on line graph and graph convolution to make the paper self-contained.

\begin{figure}
\centering
\includegraphics[scale=0.3]{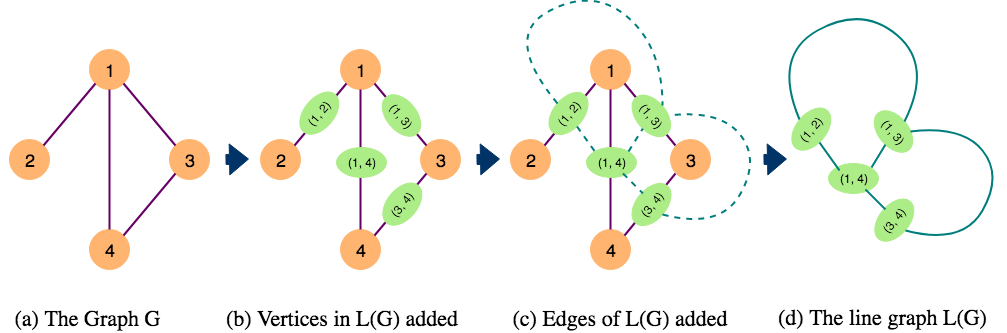}
\caption{Transformation process of a graph into its line graph.
(a) Represents a simple graph $G$. (b) Each edge in the original graph has a corresponding node in the line graph. Here the green edges represent the nodes in line graph. (c) For each adjacent pair of edges in $G$ there exists an edge in $L(G)$. The dotted lines here are the edges in the line graph. (d) The line graph $L(G)$ of the graph $G$}
\label{fig:line_graph}
\end{figure}

\subsection{Line Graph and Extension to Hypergraphs}\label{sec:LG}
First, we define a line graph of a simple graph. Given a simple undirected graph $G = (V, E)$, the line graph $L(G)$ is the graph such that each node of $L(G)$ is an edge in $G$ and two nodes of $L(G)$ are neighbors if and only if their corresponding edges in $G$ share a common endpoint vertex \cite{whitney}. Formally $L(G) = (V_{L}, E_{L})$ where $V_L = \{(v_i, v_j) : (v_i, v_j) \in E\}$ and $E_L = \{\big((v_i, v_j), (v_j, v_k)\big) : (v_i, v_j) \in E \:, \: (v_j, v_k) \in E \}$.
Figure \ref{fig:line_graph} shows how to convert a graph into the line graph. 

Now, we discuss the process of transforming 
an attributed hypergraph $\mathcal{H}=(V,E)$ (as discussed in Section \ref{sec:prob}) to an attributed weighted line graph $L(\mathcal{H}) = (V_L,E_L)$ as follows. Similar to the case of simple graph, we create a node $\mathbf{v}_e$ in the line graph for each hyperedge $e$ in the hypergraph. We connect the two nodes in the line graph by an edge if the corresponding two hyperedges in the hypergraph share at least one common node. More formally, $V_L = \{\mathbf{v}_e \;|\; e \in E\}$, and $E_L = \{\{\mathbf{v}_{e_p},\mathbf{v}_{e_q}\} \;|\; |e_p \cap e_q| \geq 1, \; e_p,e_q \in E\}$. Next, we assign a weight $w_{p,q}$ to each edge $\{\mathbf{v}_{e_p},\mathbf{v}_{e_q}\}$ of the line graph as follows.
\begin{equation}\label{eq:LGweight}
    w_{p,q} = \frac{|e_p \cap e_q|}{|e_p \cup e_q|}
\end{equation}

\begin{figure}[h!]
  \centering
    \includegraphics[width=0.8\linewidth]{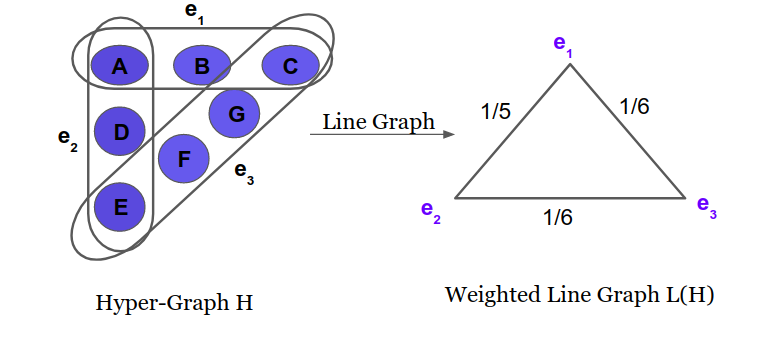}
    \caption{Transforming a hypergraph to the corresponding weighted line graph.}
    \label{fig:htoline}
\end{figure}

Example of the formulation of a line graph from a small hypergraph is depicted in Figure \ref{fig:htoline}.
The original hypergraph has attribute $x_v \in \mathbb{R}^d$ to each node $v$. We follow a simple strategy to assign attributes to each node of the corresponding line graph. For a node $\mathbf{v}_e \in V_L$, we check the corresponding hyperedge $e \in E$ of the given hypergraph. We consider each node $v \in V$ of the hypergraph which belongs to this hyperedge $e$ and we assign the attribute $\mathbf{x}_{\mathbf{v}_e} \in \mathbb{R}^d$ of $\mathbf{v}_e$ as the average attribute of all these nodes. So, $\mathbf{x}_{\mathbf{v}_e} = \frac{\sum\limits_{v \in e} x_v}{|e|}$.

Finally, as discussed in Section \ref{sec:prob}, we are interested in the semi-supervised node classification of the given hypergraph. We have subset of nodes $V^l \subset V$ in hypergraph for which we have the labels available. We again follow a simple strategy to assign labels to a subset of nodes in the hypergraph. For each node $\mathbf{v}_e \in V_L$, we put it in the set of labelled nodes $V_L^l \in V_L$ if the corresponding hyperedge $e \in E$ contains at least one labelled (hyper)node from $V^l$. For each such node $\mathbf{v}_e \in V_L^l$, the label of the node is deduced as the majority of the labels of the nodes in $e \cap V^l \subset V$. For example, in Figure \ref{fig:htoline}, if nodes in the hypergraph E and F are labelled as 1, G is labelled as 2 and C is unlabelled, then the node $e_3$ in the line graph would be labelled as 1.

\subsection{Convolution Network on the Line Graph}\label{sec:LHCN}
In this section, we aim to apply the graph convolution for the hypergraphs. Please note, the proposed line graph of a hypergraph is completely different from the concept of dual graph of a hypergraph \cite{berge1973graphs}. Dual graphs are formed just by interchanging vertices and edges of a hypergraph. The dual graph of a hypergraph is still a hypergraph. Whereas from the last section, one can see that, even though the input is a hypergraph, the resulting line graph is a simple graph (i.e., each edge connects only two nodes). Also, we derive attributes to each node of the line graph and there are positive weight associated with each edge. Hence, this enable us to apply graph convolution on the generated line graph, as explained below.

For the deduced line graph $L(H)$, say $A \in \mathbb{R}^m$ is the adjacency matrix of it, where $a_{pq} = w_pq$ (as mentioned in Equation \ref{eq:LGweight}). To apply the graph convolution, we add self loops to each node of the line graph and normalize it. Say, $\hat{A}=A+I$, where $I$ is an identity matrix. $\hat{D}$ is a diagonal matrix with $\hat{D}_{p,p} = \sum\limits_{q=1}^m \hat{A}_{p,q}$. For all our experiment, we use the following 2-layered graph convolution on the line graph:
\begin{equation}\label{eq:LGNR}
    H = \sigma(\hat{D}^{-\frac{1}{2}} \hat{A} \hat{D}^{-\frac{1}{2}} \underbrace{\sigma(\hat{D}^{-\frac{1}{2}} \hat{A} \hat{D}^{-\frac{1}{2}} \mathbf{X} \Theta^{(1)})}_\text{Intermediate representation} \Theta^{(2)} )
\end{equation}

Here, $\hat{D}^{-\frac{1}{2}} \hat{A} \hat{D}^{-\frac{1}{2}}$ is the symmetric graph normalization, $\sigma()$ is a nonlinear activation function, for which we use Leaky Relu. $\Theta^{(1)} \in \mathbb{R}^{d \times k_1}$ and $\Theta^{(2)} \in \mathbb{R}^{k_1 \times k}$ are the learn-able weight parameters of GCN which transforms the input feature matrix $\mathbf{X}$ and the intermediate hidden representations respectively. $H \in \mathbb{R}^{m \times k}$ is the final hidden representation of the nodes of the line graph, which are fed to a softmax layer and use the following cross entropy loss for node classification.
\begin{equation}
    \mathcal{L}oss = - \sum\limits_{p \in V_L^l} \sum\limits_{l \in \mathcal{L}} y_{p,l} \; \ln{\hat{y}_{p,l}}
\end{equation}
Here, $y_{p,l}$ is an indicator function if the actual label of a node $p$ in $V_L^l$ of the line graph is $l$, as derived in Section \ref{sec:LG}, and $\hat{y}_{p,l}$ is the probability with which the node $p$ has the label $l$ from the output of softmax function. We use back-propagation algorithm with ADAM optimization technique \cite{kingma2014adam} to learn the parameters of the graph convolution on the cross entropy loss. After the training is completed, we obtain the labels of all the nodes in the line graph.

\subsection{Information Propagation to the Hypergraph}
Here, we move the learned node labels and the representations back from the line graph $L(\mathcal{H})$ to the given hypergraph $\mathcal{H}$. Each node in the line graph corresponds to a hyperedge of the hypergraph. So after running the graph convolution in Section \ref{sec:LHCN}, we obtain labels and representation of each hyperedge. We again opted for a simple strategy to get them for the nodes of the hypergraph.
For an unlabelled node $v \in V^u \subset V$, we take the majority of the labels of the edges it belongs to. For example, if the node $v$ is part of three hyperedges $e_1$, $e_2$ and $e_3$, and if their labels are 1, 2 and 2, then the labels of the node $v$ is assigned as 2. Similarly to get the vector representation $h_v \in \mathbb{R}^k$ of a node $v$ in the hypergraph, we take the average representations of all the hyperedges it belongs to.

\begin{figure*}[thb]
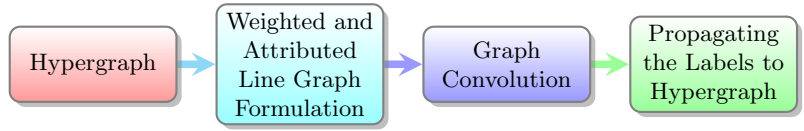

    \centering
    \smartdiagramset{
        text width=2cm,
        back arrow disabled=true}
    \smartdiagram[flow diagram:horizontal]{Hypergraph, Weighted and Attributed Line Graph Formulation, Graph Convolution, Propagating the Labels to Hypergraph}
    \caption{Different stages of LHCN algorithm}
    \label{fig:LHCN}
\end{figure*}

Though the above scheme to generate hypernode representations from the node representations (as in Equation \ref{eq:LGNR}) of the line graph is simple in nature, it completely adheres the hypergraph Laplacian assumption that any two nodes connected by an hyperedge tend to have similar embeddings and labels. This also satisfies the well-known concept of homophily \cite{mcpherson2001birds} in social networks. The different stages of LHCN are summarized in the block diagram in Figure \ref{fig:LHCN}.

\textbf{Time Complexity of LHCN}: As discussed in Section \ref{sec:prob}, the given hypergraph has $n$ number of (hyper)nodes and $m$ hyperedges. We assume feature dimension $d$ and embedding dimension $k$ are constants. On average, if a node belong to $s$ number of hyperedges, all the corresponding hyperedges will be connected to each other in the line graph. Hence, complexity of generating the line graph is $O(m+ns^2)$. The runtime of graph convolution in the line graph would take $O(ns^2)$ time. Propagating the information back to the hypergraph requires $O(m)$ time. Hence, the total runtime of LHCN is $O(m+ns^2)$. For real work sparse graphs, $s$ is a constant. Hence the effective runtime of LHCN is $O(m+n)$.


\section{Experimental Evaluation}\label{sec:exp}
We experiment on three publicly available popular network datasets for node classification and representation and compare the results with state-of-the-art hypergraph neural networks. Please note, as the inherent objective of graph convolution is semi-supervised in nature, we do not conduct experiment on unsupervised tasks such as hypernode clustering and hyperedge prediction. This is consistent with the hypergraph neural network literature \cite{yadati2019hypergcn,feng2019hypergraph}.

\subsection{Datasets and Baseline Algorithms Used}
We use three citation network datasets: Cora, Citeseer and Pubmed. We construct hypergraphs from them, as done in recent hypergraph neural network papers \cite{yadati2019hypergcn,feng2019hypergraph}. We keep all the nodes of the original network in the hypergraph. If a research paper `a' cites the papers `b',`c' and `d', then we create a hyperedge $\{a,b,c,d\}$ in the hypergraph. Each node in these datasets is associated with an attribute vector. Attributes represent the occurrence of a word in the research paper through bag-of-words models. For Cora and Citeseer the attributes are binary vector and for Pubmed, they are tf-idf vectors. High level summary of these datasets are presented in Table \ref{tab:datasets}.

\begin{table}[]
    \centering
    \setlength{\tabcolsep}{15pt}
    \begin{tabular}{*4c}
    \toprule
	\textbf{} & \textbf{Cora}  & \textbf{Citeseer } & \textbf{Pubmed} \\
	\hline
	\midrule
	Number of hypernodes  & 2708 & 3312 & 19717 \\
	Number of hyperedges  & 1579 & 1079 & 7963 \\
	Number of features  & 1433 & 3703 & 500 \\
	Number of Classes  & 7 & 6 & 3 \\
    \bottomrule
    \end{tabular}\\
    \caption{Citation hypergraph datasets used in this work.}
    \label{tab:datasets}
\end{table}

We have selected the following set of diverse state-of-the-art algorithms to compare with the results of the proposed algorithm LHCN.
\begin{itemize}
    \item \textbf{Confidence Interval based method (CI)} \cite{zhang2017re}: Authors have proposed a semi-supervised learning approach on hypergraphs and design an algorithm for solving the convex program based on the subgradient method.
    \item \textbf{Multi-layer perceptron (MLP)}: This is a classical MLP which only uses the node attributes for node classification problem. The hypergraph structure is completely ignored here.
    \item \textbf{MLP + explicit hypergraph Laplacian regularisation (MLP + HLR)}: Here the MLP is used along with an added component to capture the hypergraph Laplacian regularizer \cite{yadati2019hypergcn}.
    \item \textbf{Hypergraph Neural Networks}: Here we used recently proposed state-of-the-art hypergraph neural networks. First, \textbf{HGNN} \cite{feng2019hypergraph} is considered as it uses graph convolution on the weighted clique expansion of a hypergraph. Second, we consider different variants of \textbf{HyperGCN} as explained in \cite{yadati2019hypergcn}. HyperGCN approximates hypergraph Laplacian into a simple graph and then uses graph convolution in it to classify the nodes of the hypergraph.
\end{itemize}

\begin{table}[]
    \centering
    \setlength{\tabcolsep}{15pt}
    \begin{tabular}{*4c}
    \toprule
    \\
	\textbf{Method} & \textbf{Cora}  & \textbf{Citeseer } & \textbf{Pubmed} \\
	\\
	\hline
	\midrule
	CI & 35.60 $\pm$ 0.8 & 29.63 $\pm$ 0.3  &47.04 $\pm$ 0.8 \\
    MLP &57.86 $\pm$ 1.8 & 58.88 $\pm$ 1.7 & 69.30 $\pm$ 1.6 \\
    
    MLP+HLR &63.02 $\pm$ 1.8 & 62.25 $\pm$ 1.6 & 69.82 $\pm$ 1.5  \\
    \hline
    HGNN & 67.59 $\pm$ 1.8 & 62.60 $\pm$ 1.6 & 70.59 $\pm$ 1.5  \\
    1-HyperGCN & 65.55 $\pm$ 2.1 & 61.13 $\pm$ 1.9 & 69.92 $\pm$ 1.5  \\
    FastHyperGCN & 67.57 $\pm$ 1.8  & 62.58 $\pm$ 1.7 & 70.52 $\pm$ 1.6  \\
    HyperGCN & 67.63 $\pm$ 1.7 & 62.65 $\pm$ 1.6 & \textbf{74.44} $\pm$ \textbf{1.6}  \\
    \hline
    \textbf{LHCN} (ours) & \textbf{73.34} $\pm$ \textbf{1.7} & \textbf{63.19} $\pm$ \textbf{2.23} & 70.76 $\pm$ 2.36  \\
    \bottomrule
    \end{tabular}\\
    \caption{Results of node classification (accuracy with standard deviation  in \%).}
    \label{tab:nodeClassi}
\end{table}

\subsection{Experimental Setup}
For all our experiments, we split the (hyper)nodes into 80-20\% random train-test split and then train LHCN (with a two layer GCN on the generated line graph). We decrease the learning rate of the ADAM optimizer after every 100 epochs by a factor of 2. For Cora and Citeseer datasets, the dimensions of the first and second hidden layers of GCN are 32 and 16 respectively and we train the GCN model for 200 epochs. Pubmed being a larger dataset, the dimesnions of the hidden layers are set to 128 and 64 respectively, and we train it for 1700 epochs.
We perform the experiments in a shared server having Intel(R) Xeon(R) Gold 6142 processor which contains 64 processors with 16 core each. The overall runtime (including all the stages) of LHCN on Cora, Citeseer and Pubmed datasets are approximately 16 sec., 21 sec. and 867 sec. respectively.

\subsection{Results on Node Classification}
We run LHCN (on 80-20\% random train test splits) 10 times on each dataset and reported the average node classification accuracy and standard deviation in Table \ref{tab:nodeClassi}. Reported numbers for the baselines are taken from \cite{yadati2019hypergcn}. First, it can be observed that graph neural network based approaches perform better than the rest. Our proposed LHCN improves the performance on Cora and Citeseer datasets. On Cora dataset, LHCN is able to outperform the closest baseline HyperGCN by roughly 8.4\%. Very recently proposed HyperGCN turns out to be the best performer on Pubmed, and many of the baseline algorithms, including ours are able to generate accuracy around 70\%.

\subsection{Hypernode Visualization}
\begin{figure*}[h!]
  \centering
  \begin{subfigure}[b]{0.32\linewidth}
    \includegraphics[width=\linewidth]{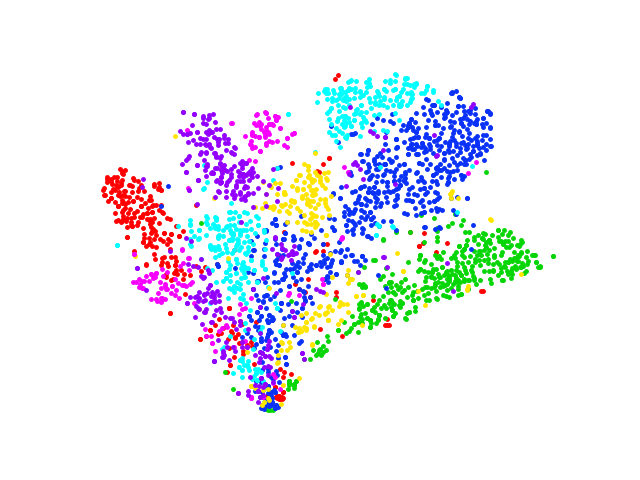}
    \caption{\scriptsize Cora}
    \label{fig:cora_tsne}
  \end{subfigure}
  \begin{subfigure}[b]{0.32\linewidth}
    \includegraphics[width=\linewidth]{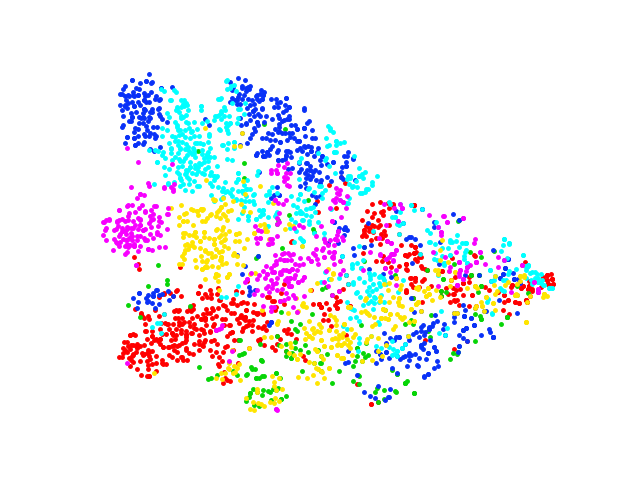}
    \caption{\scriptsize Citeseer }
    \label{fig:viz-IIL}
  \end{subfigure}
  \begin{subfigure}[b]{0.32\linewidth}
    \includegraphics[width=\linewidth]{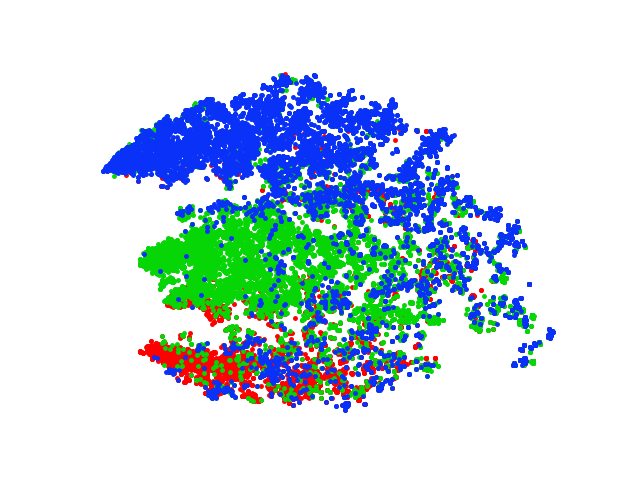}
    \caption{\scriptsize Pubmed}
    \label{fig:pubmed_cora}
  \end{subfigure}
  \caption{t-SNE visualization of the hypernode representations generated by LHCN}
  \label{fig:tsne}
\end{figure*}

LHCN produces vector representations of the (hyper)nodes as well. To see the visual quality of those representations, we t-SNE \cite{vanDerMaaten2008} which maps the vectors to two dimensional space which can be plotted easily. Figure \ref{fig:tsne} shows the visualization of hypernode representations of all the datasets using LHCN. Different colors represent different node classes. We can observe that the classes are separated to a good extent for all the datasets. We also tried to run HyperGCN to generate the hypernode representations. But the visual quality of the representations are not coming good. So, we present the results only for our algorithm.

\subsection{Further Analysis on LHCN}
\begin{figure*}[h!]
  \centering
  \begin{subfigure}[b]{0.46\linewidth}
    \includegraphics[width=\linewidth]{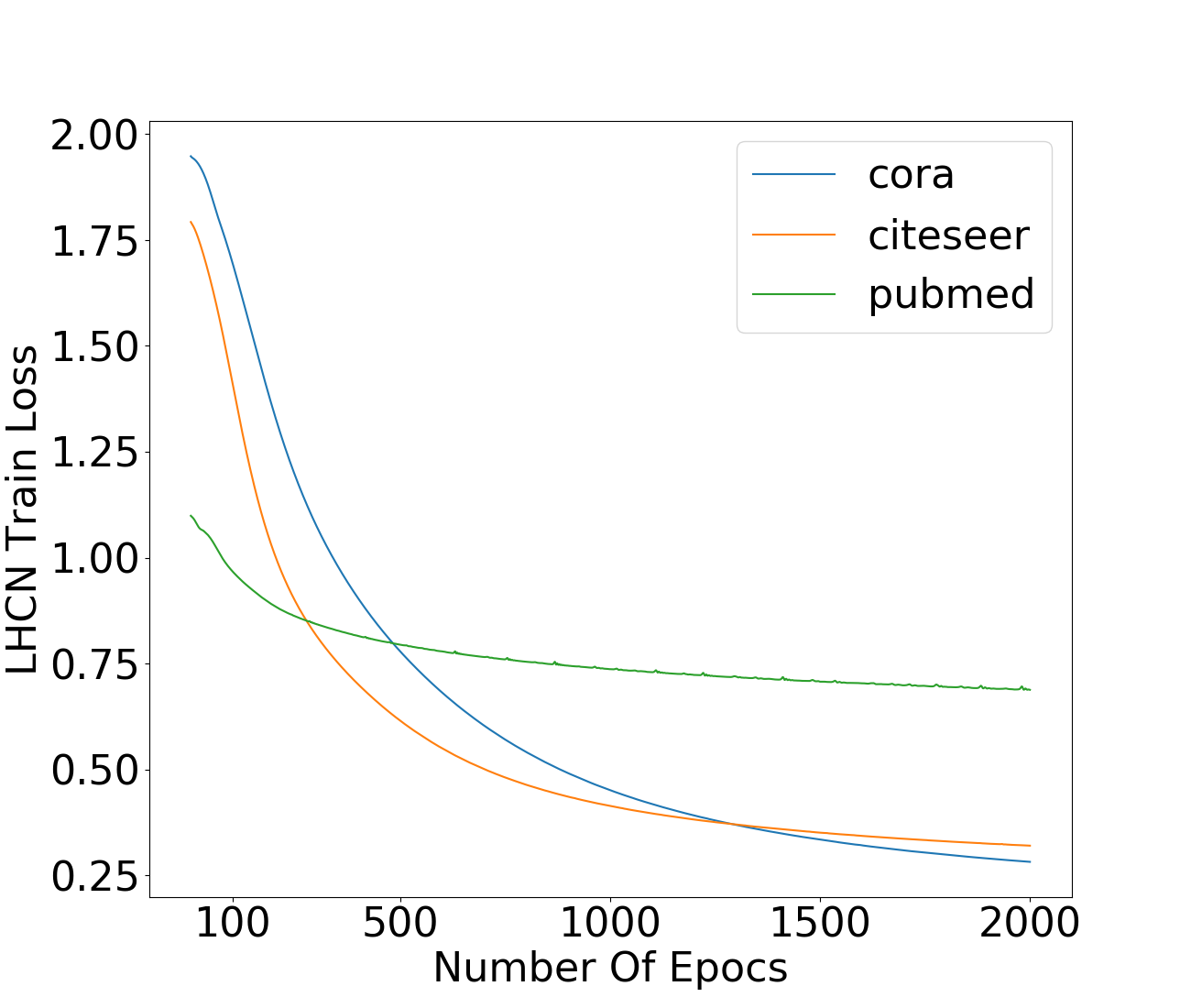}
    \caption{}
    \label{fig:loss}
  \end{subfigure}
  \begin{subfigure}[b]{0.46\linewidth}
    \includegraphics[width=\linewidth]{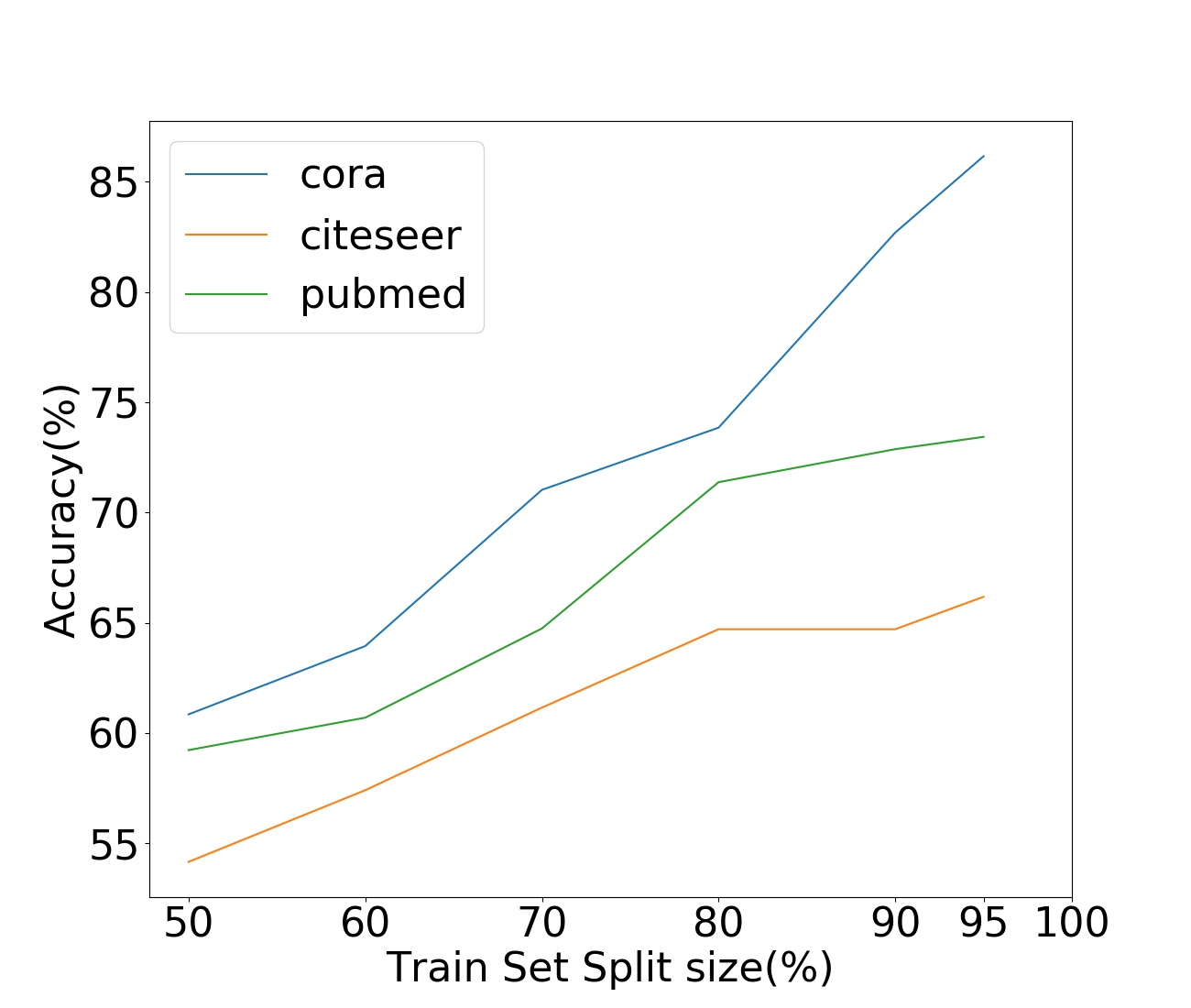}
    \caption{\scriptsize}
    \label{fig:accu_vs_split}
  \end{subfigure}
  \caption{(a) shows LHCN training Loss over different epochs of the algorithm. (b) shows node classification accuracy over different train-test sizes.}
\end{figure*}

We conduct few more experiments to get more insight about LHCN. First, we observe the convergence of training loss over the epochs of LHCN in Figure \ref{fig:loss}. Due to the use of ADAM and our strategy to decrease the learning rate after every 100 epochs, the training losses are decreasing fast at the beginning and converge at the end.
In Figure \ref{fig:accu_vs_split}, we change the training size from 50\% to 90\% with 10\% increment. At the end, we include the case when training size is 95\% and remaining is the test set. We can see, the mode classification performance of LHCN improves significantly over increasing training set. This implies that the algorithm is able to use the labelled nodes properly to improve the results.

\section{Conclusion and Future Work}\label{sec:con}
In this work, we propose a novel approach of applying graph convolution to a hypergraph via a transformation to a weighted and attributed line graph. Experimental results are promising and improve state-of-the-arts in this recently developed area of graph neural networks for hypergraphs. There is ample scope of future work in this direction. One practical issue is the suitable construction of the hypergraph so that algorithms can extract meaningful information easily. In our work, we adopt a rule to transform a hypergraph to a line graph. It would be interesting to study if this rule can be learned from the data.  

%
%
\bibliographystyle{splncs04}
\bibliography{LHCN}

\end{document}